\documentclass[aps,twocolumn,superscriptaddress,showpacs,amssymb,prb,floatfix,longbibliography]{revtex4-2}
\usepackage{graphicx }
\usepackage[]{hyperref}
\usepackage{amsmath}
\usepackage{amssymb}

\hypersetup{colorlinks=true,linkcolor=blue,citecolor=blue,urlcolor=blue,pdfpagemode=UseNone}

\newcommand{\slrrtext}  {spin-lattice-relaxation rate}
\newcommand{\slrr}      {$T_1^{-1}$}
\newcommand{\tmvo}      {TmVO$_4$}

\begin{document}

\title{Anisotropic nematic fluctuations above the ferroquadrupolar transition in TmVO$_4$}
\author{Z. Wang}
\author{I. Vinograd}
\author{Z. Mei}
\affiliation{Department of Physics, University of California, Davis, California
95616, USA}
\author{P. Menegasso}
\affiliation{Instituto de F\'isica \lq\lq Gleb Wataghin\rq\rq, UNICAMP, Campinas-SP, 13083-859, Brazil}
\author{D. Garcia}
\affiliation{Department of Physics, University of California, Davis, California
95616, USA}\author{P. Massat}
\author{I. R. Fisher}
\affiliation{Geballe Laboratory for Advanced Materials and Department of Applied Physics, Stanford University,  CA 94305, USA}
\author{N. J. Curro}
\affiliation{Department of Physics, University of California, Davis, California
95616, USA}
\date{\today}
\begin{abstract}

TmVO$_4$ exhibits ferroquadrupolar order below 2.15 K with a well-isolated non-Kramers ground state doublet, and is a model system to understand Ising nematic order.  We present $^{51}$V nuclear magnetic resonance data as a function of field orientation in a single crystal.  Although the spectra are well understood in terms of direct dipolar hyperfine couplings, the spin lattice relaxation rate exhibits strong anisotropy that cannot be understood in terms of magnetic fluctuations.  We find that the spin lattice relaxation rate scales with the shear elastic constant associated with the ferroquadrupole  phase transition, suggesting that quadrupole (nematic) fluctuations dominate the spin lattice relaxation for in-plane fields.

\end{abstract}


\maketitle

\section{Introduction}

Recently there has been growing interest in the behavior of electronic Ising nematicity, which may play a role in the low temperature behavior of a number of strongly correlated electron systems.  The phase diagram of the iron-based superconductors is dominated by $C_4$ symmetry breaking of the spin and orbital degrees of freedom, 
accompanied by a tetragonal-to-orthorhombic structural transition \cite{FernandesSchmalianNatPhys2014}. 
The physical origin of the nematicity is yet to be established, and might even be different in different families of compounds, with possible contributions both from orbital effects and spin fluctuations. The correlation between large values of the nematic susceptibility, a putative nematic quantum critical point, and optimal superconductivity in several materials, point to a possible role for nematic fluctuations in the pairing interaction, as well as non-Fermi-liquid behavior in the normal state \cite{FisherScienceNematic2012,Kuo2015,KivelsonNematicQCP2015,ScalapinoNematicQCP2015}.  
Electronic nematic correlations are also present in the high temperature superconducting cuprates \cite{Kivelson1998,Vojta2009}. Disentangling the effects of nematicity can be complicated by the presence of other intertwined order parameters in such systems. It is challenging to discern whether nematic or antiferromagnetic fluctuations dominate, especially in a conducting system \cite{DioguardiPdoped2015}.  Moreover, strain couples to the nematic order, hence quenched random disorder around dopants may give rise to inhomogeneous glassy behavior \cite{DioguardiNematicGlass2015,HysteresisNematicPRL}. It is therefore important to investigate critical fluctuations in an Ising nematic system in the absence of metallicity or inhomogeneous strain fields.

\begin{figure}
\centering
\includegraphics[width=\linewidth]{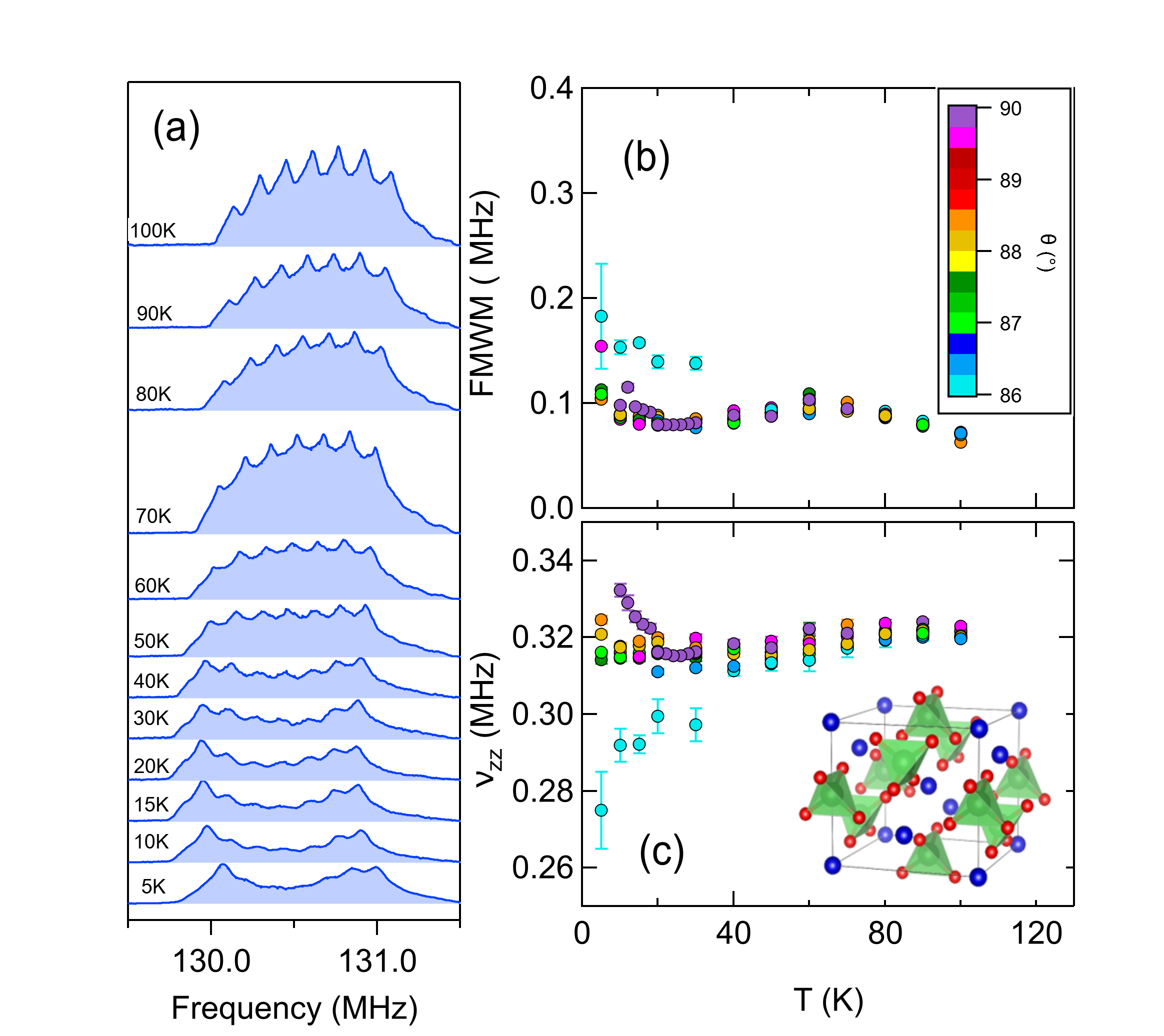}
\caption{\label{Fig:spectra} (a) $^{51}$V spectra in TmVO$_4$ at 11.7294T for $\mathbf{H}_0\perp c$ at several different temperatures.  The full-width half maximum (b) and quadrupolar splitting (c) versus temperature for several different angles, $\theta$. The inset shows the unit cell structure.  Tm is blue, V is green (within the pyramids) and O is red.}
\end{figure}

Ferroquadrupolar ordering of non-Kramers doublets in 4f materials offers an important avenue to investigate Ising nematicity \cite{FisherNematicQCP}. Specifically, ferroquadrupole order breaks all the same symmetries as Ising nematic order (and hence is a specific realization of nematic order), with the advantage that the underlying effective Hamiltonian describing the low-temperature behavior is well-understood. \tmvo\ is an insulator with Tm ions in the 4f$^{12}$ configuration  ($L=5$, $S=1$, $J=6$) that crystallizes in space group $I41/amd$ (see inset of Fig. \ref{Fig:spectra}). The tetragonal crystal field splits the $J=6$ multiplet giving rise to a $\Gamma_5$ ground state doublet separated by a gap of $\sim 77$ K to the lowest excited state \cite{BleaneyTmVO4review,Bowden1998}.  The wavefunctions of the ground state doublet are $|\psi_{1,2}\rangle = e|\pm 5\rangle + f|\pm 1\rangle + g|\mp 3\rangle$ in the $J_z$ basis, where $e\approx 0.92$, $f\approx -0.37$, and $g\approx 0.12$.  The degeneracy of these ground states cannot be lifted by a magnetic field perpendicular to the $z$-direction because $\langle \psi_{1, 2}|J_{\pm}|\psi_{1, 2}\rangle = 0$, hence these form a non-Kramers doublet. 
The doublet can, however, be linearly split by either a magnetic field oriented along the $c$-axis, or by lattice strains with either a $B_{1g}$ ($x^2-y^2$) or $B_{2g}$ ($xy$) symmetry. Quadrupole-quadrupole interactions mediated by the lattice dominate the magnetic interactions, and the material undergoes
a cooperative Jahn-Teller ferroquadrupolar ordering with a  $B_{2g}$ symmetry at $T_Q = 2.15$K, accompanied by an orthorhombic lattice distortion of the same symmetry \cite{Gehring1975}.  The low-temperature behavior of the Tm quadrupoles can be well-described by the transverse field Ising model, in which  pseudospins ($\tilde{S}={1}/{2}$) experience in-plane ferroquadrupolar Ising couplings 
and couple to a transverse magnetic field along the $z$-axis \cite{FisherNematicQCP}. These fields will enhance the fluctuations of the pseudospins and can tune the system to an Ising-nematic quantum phase transition. This material thus offers an important platform to investigate quantum critical nematic fluctuations in an insulator.

In order to better understand the nature of these fluctuations we have investigated $^{51}$V ($I=7/2$, $Q=52$ mb, 99.75\% abundant) NMR in a single crystal of \tmvo\ in a magnetic field $\mathbf{H}_0 = 11.72$ T oriented perpendicular to the $c$-axis.  We find that the magnetic shift and \slrrtext\ are strongly angular dependent (rotating the field in the $a-c$ plane)  below $\sim 80$ K, reflecting the anisotropy of the ground state doublet.  The \slrrtext\ is non-monotonic, exhibiting a large enhancement at low temperature that may be associated with critical nematic fluctuations.  The spectrum exhibits a strong temperature-dependent magnetic broadening at low temperatures caused by inhomogeneous magnetic demagnetization fields.     

\begin{figure}
\centering
\includegraphics[width=\linewidth]{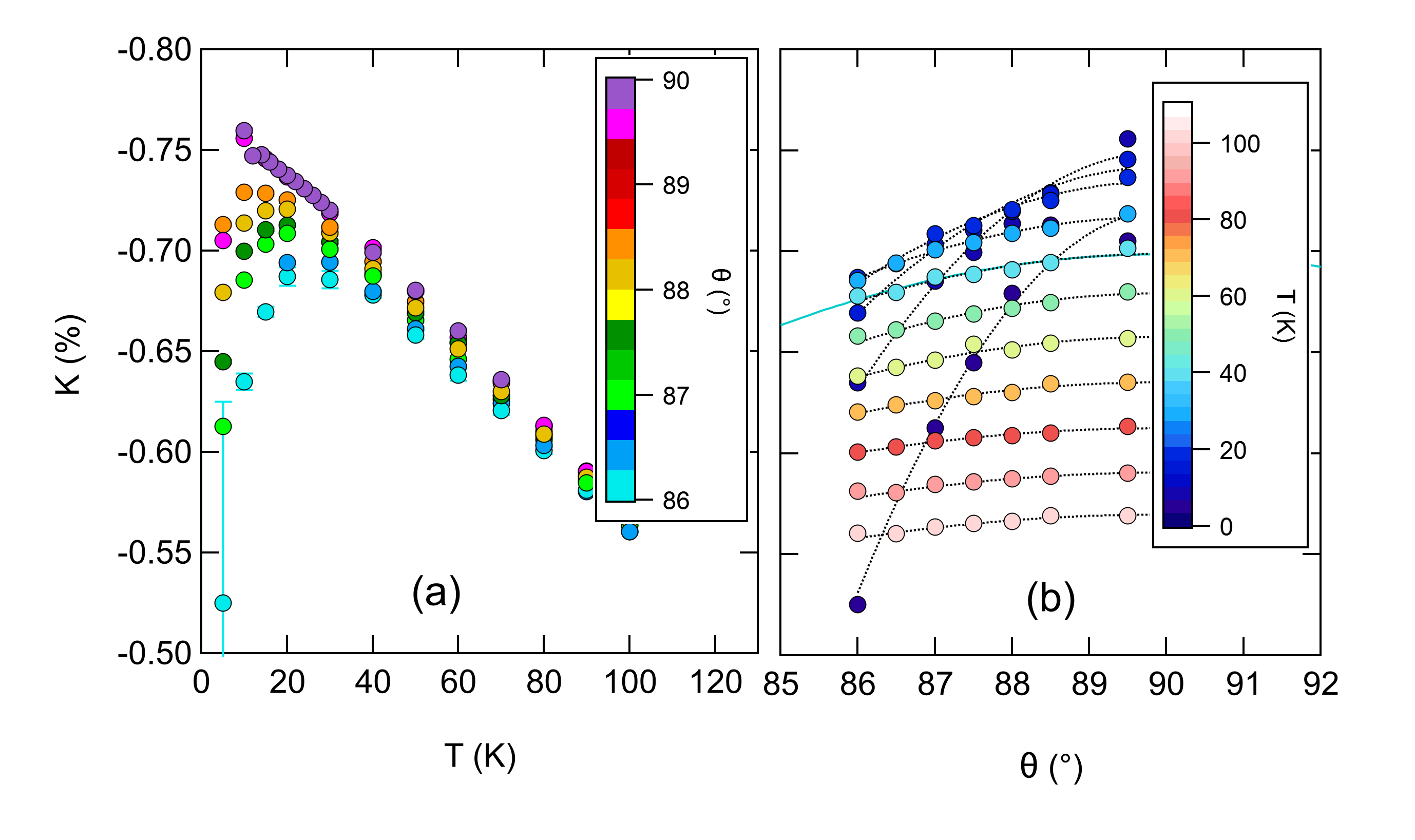}
\caption{\label{Fig:shift} magnetic shift, $K$, versus temperature (a) and versus angle (b). The dotted lines are fits as described in the text.}
\end{figure}

\section{NMR Spectra and magnetic shift}

\tmvo\ crystals were grown from a Pb$_2$V$_2$O$_7$ flux using 4 mole percent of Tm$_2$O$_3$, following the methods described in \cite{FEIGELSON1968,Smith1974}. The  crystals have a rod-like morphology with the the $c$-axis along the long axis. A single crystal of approximate dimension 1mm $\times$ 1mm $\times$ 4mm was selected and mounted on a cryogenic goniometer NMR probe. The magnetic susceptibility is strongly anisotropic, reflecting that of the unusual  $g$-factor ($g_c= 10.2$, $g_{\perp} ~ 0$) of the ground state doublet.
Although a crystal mounted with $c\perp\mathbf{H}_0$ experiences zero torque, it is an unstable equilibrium and there is a large torque for infinitesimal deviations from 90$^{\circ}$. To alleviate this issue we secured the crystal with epoxy to a mounting plate that itself is rotated. Spin echoes were acquired at several different frequencies, and the Fourier transforms were summed to measure the full spectra including all nuclear spin transitions.  Fig. \ref{Fig:spectra}(a) shows several representative spectra of the $^{51}$V  as a function of temperature. There are seven peaks separated by the quadrupolar interaction.  Because the V has axial symmetry, the peaks frequencies are given by: 
\begin{equation}
\nu = \gamma H_0\left( 1 + K(\theta)\right) + n\nu_{q}(\theta),    
\end{equation}
where the magnetic, $K(\theta)$, and quadrupolar, $\nu_{q}(\theta)$, shifts vary with the angle $\theta$ between the c-axis and $\mathbf{H}_0$:
\begin{eqnarray}
K(\theta) &=&K_{cc}\cos^2\theta + K_{aa}\sin^2\theta\\
\nu_{q}(\theta) &=& \nu_{zz}\left(3\cos^2\theta - 1\right)/2.
\end{eqnarray}
Here $\gamma = 11.193$ MHz/T, $n = -3, \cdots, 3$, $\nu_{zz}={eQV_{zz}}/{12 h}$ and  $V_{zz}$ is the electric field gradient at the V site. We measured the spectra for several angles $86^{\circ}\leq\theta\leq 90^{\circ}$ and fit the spectra to a sum of Lorentzians.  The temperature and angular dependence of the linewidths, EFG, and magnetic shifts are shown in Figs. \ref{Fig:spectra}(b, c) and \ref{Fig:shift}(a).  

We find that the EFG is similar to previous measurements, \cite{BleaneyTmVO4review,Bleaney1983} 
however the spectra are broad.  For this orientation, the quadrupolar splitting is $\nu_{zz}/2$, which is comparable to the FWHM of each resonance.  As a result, the individual peaks become difficult to resolve at low temperatures.  Each of the satellite resonances has the same linewidth, implying that that the broadening is due to magnetic field inhomogeneity within the sample.   Moreover, we find that the spectra are narrower at lower applied fields and that the FWHM varies approximately linear with field.  Since the susceptibility is strongly anisotropic, distortions of the internal field $\mathbf{B}$ due to demagnetization effects within the needle-like prism of the crystal can create a large distribution of local resonance frequencies \cite{Lawson2018}.  The spectra also display a suppression of intensity for the inner satellites, particularly at low temperature.  This phenomenon arises due to fast spin-spin decoherence rates ($T_{2}^{-1}$) with high spin nuclei \cite{NissonBi2Se3NMR}.

\begin{figure}
\centering
\includegraphics[width=\linewidth]{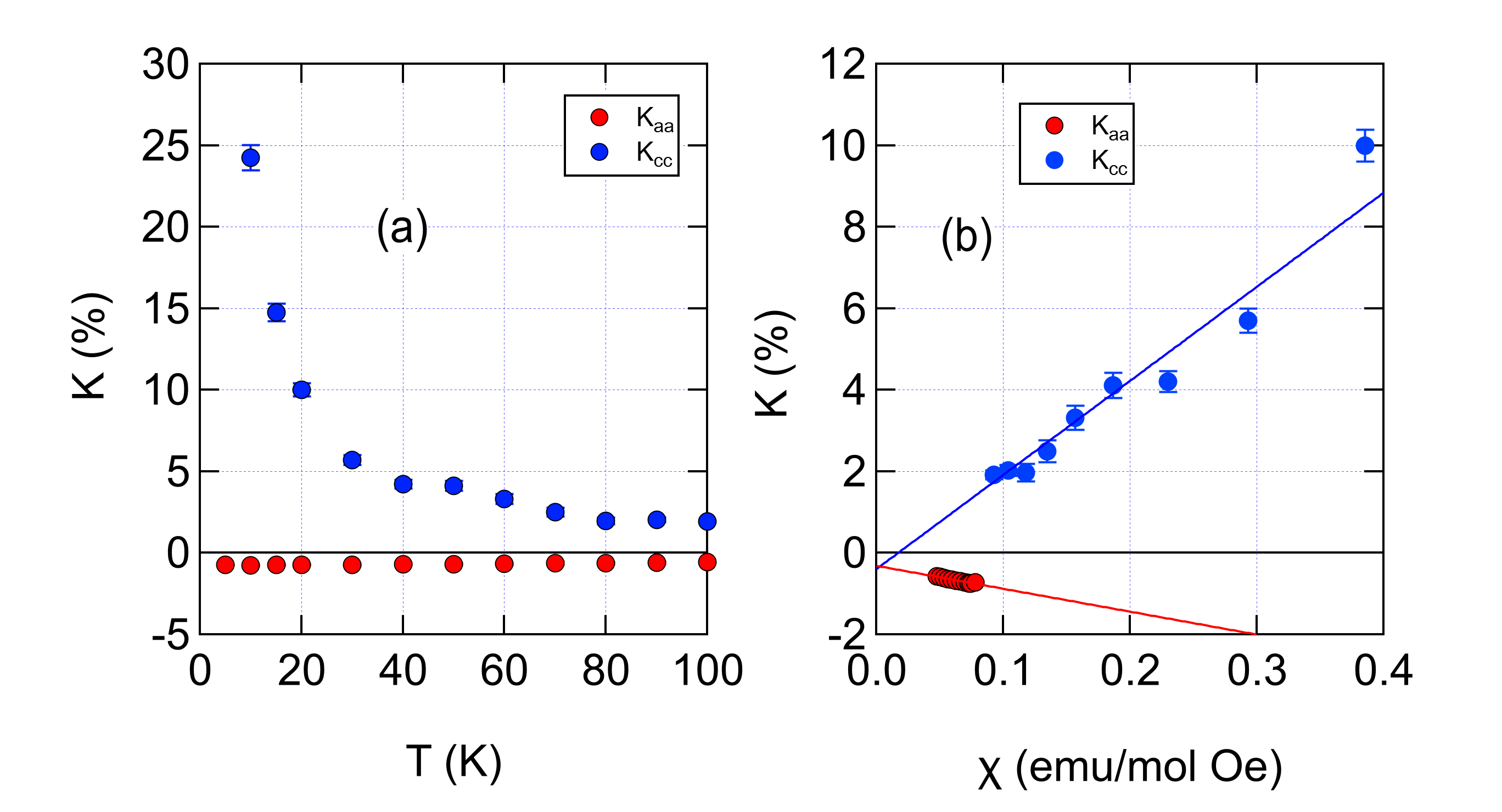}
\caption{\label{Fig:Kani} $K_{aa}$ and $K_{cc}$ versus temperature (a) and versus bulk susceptibility (b).  The solid lines are fits as described in the text.}
\end{figure}

The magnetic shift shown in Fig. \ref{Fig:shift} is negative and strongly angular dependent at low temperature.  We fit the angular dependence to extract the tensor components $K_{aa}$ and $K_{cc}$, shown as dotted lines in Fig. \ref{Fig:shift}(b).   This approach enables us to extract the magnetic shift for the $c$ direction without needing to fully align the crystal in this orientation, albeit the error bars for $K_{cc}$ are larger than for $K_{aa}$.   The temperature dependence of $K_{aa}$ and $K_{cc}$ are shown in Fig. \ref{Fig:Kani}(a). $K_{cc}$ is large and positive. Fig. \ref{Fig:Kani}(b) shows these shift components plotted versus the bulk susceptibility $\chi_{aa, cc}$, which was measured independently in a SQUID magnetometer.  The shift varies linearly with susceptibility as $K_{\alpha\alpha} = K_{\alpha\alpha}^{orb} + A_{\alpha\alpha}\chi_{\alpha\alpha}$, where $A_{\alpha\beta}$ are the components of the hyperfine coupling tensor.  We find that $K_{aa}^{orb} =-0.315\pm0.009\%$, $K_{cc}^{orb} =-0.4\pm0.1\%$, $A_{aa} =-0.32\pm0.07$ kOe/$\mu_B$ and $A_{cc} =1.29\pm0.05$ kOe/$\mu_B$. 

These values of the hyperfine couplings are consistent with a direct dipolar coupling mechanism between the Tm moments and the V nuclear spins.  The direct dipolar coupling is given by:
$A_{\alpha\beta}^{dip} = \sum_i(\nabla\times \mathbf{A}_i)_{\alpha}/\mu_{\beta}$, where $\mathbf{A}_i = \mathbf{\mu}\times\mathbf{r}_i/r_i^3$ is the vector potential of a dipole moment, $\mathbf{\mu}$, located at lattice site $\mathbf{r}_i$ relative to a central nucleus. For the \tmvo\ lattice, we estimate $A_{aa}^{dip} = A_{bb}^{dip} = -0.336$ kOe/$\mu_B$ and $A_{cc}^{dip} = 0.671$ kOe/$\mu_B$ at the V site. The theoretical value for the perpendicular direction is the same as the measured value within the error limits.  For the $c$ axis, the theoretical value is within a factor of two of the measured values, and it is likely there are larger systematic measurement errors involved in extracting this value.  Thus the anisotropic magnetic shift tensor can be fully explained via direct dipolar interactions, as expected for an insulator.   

\section{Spin Lattice Relaxation Rate}

\begin{figure}
\centering
\includegraphics[width=\linewidth]{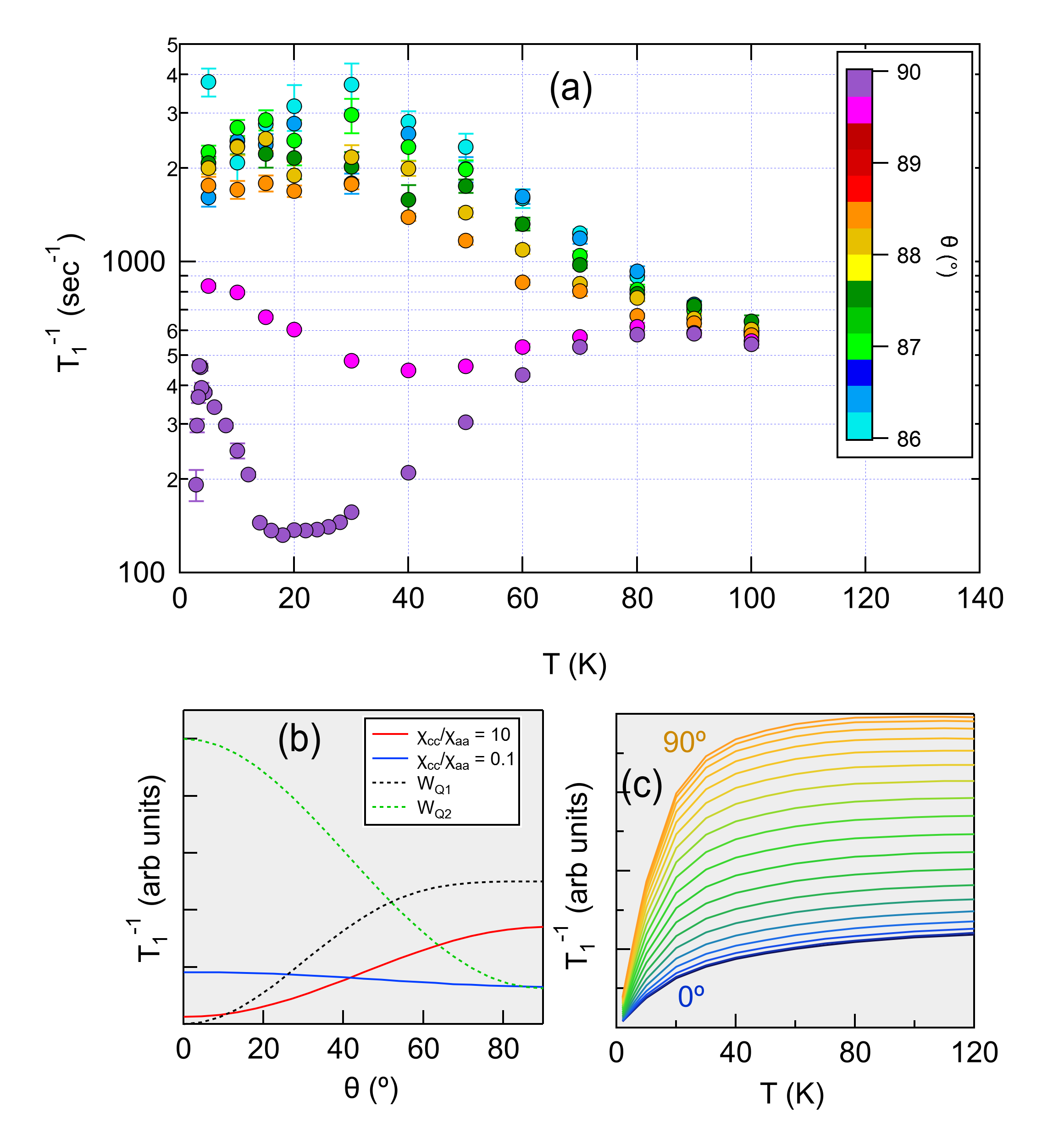}
\caption{\label{Fig:T1inv} (a) \slrr\ vs temperature for multiple angles. (b) Calculated \slrr\ versus $\theta$ for magnetic fluctuations (solid lines) and for quadrupole fluctuations (dashed lines). (c) Calculated \slrr\ versus temperature and angle using Eq. \ref{eqn:moriya}, where the angle is increased in $5^{\circ}$ increments between 0 and 90$^{\circ}$.}
\end{figure}

The \slrrtext, \slrr, was measured by applying inversion pulses at the central transition $(n=0)$ and measuring the echo intensity as a function of recovery time.  The magnetization recovery was fit to the standard expression for magnetic fluctuations: $M(t) = M_0\left(1-2f \phi(t/T_1)\right)$, where $M_0$ is the equilibrium magnetization, $f$ is the inversion fraction, and 
\begin{equation}
\label{eqn:relax}
    \phi(t) =\frac{1225}{1716}e^{-28t}+ \frac{75}{364}e^{-15t} + \frac{3}{44}e^{-6t} + \frac{1}{84}e^{-t}.
\end{equation}
This expression fits the data well without the need for a stretching exponent. Fig. \ref{Fig:T1inv} shows the temperature and angular dependence of \slrr. For $\theta = 90^{\circ}$, \slrr\ decreases strongly below 80K as the excited crystal field levels are thermally depopulated. In this temperature range \slrr\ become strongly angular dependent, increasing by more than a factor of 30 as the field $\mathbf{H}_0$ rotates by only 4$^{\circ}$ away from the perpendicular configuration.  
This behavior likely reflects the anisotropy of the $g$ factor of the ground state doublet, however the anisotropy of \slrr\ is puzzling.  If the relaxation is driven by magnetic fluctuations of the Tm ground state, then \slrr\ should exhibit a \emph{maximum} at $\theta=90^{\circ}$ rather than a minimum because fluctuations of the non-Kramers doublet should lie exclusively along the $c$-axis. Therefore $T_1^{-1}(0^{\circ})$ should be much smaller than $T_1^{-1}(90^{\circ})$, in contrast to our observations.

\subsection{Magnetic fluctuations}

On the other hand, the hyperfine couplings can give rise to a more complicated relationship between the direction of the Tm moments and the direction of the hyperfine fields.  To properly account for these couplings we use the Moriya expression:
\begin{equation}
	\label{eqn:moriya}
	T_{1m}^{-1} = \gamma^2 k_B T \lim_{\omega \rightarrow 0} \sum\limits_{\mathbf{q},\alpha,\beta} \mathcal{F}_{\alpha\beta}(\mathbf{q}) \frac{\textrm{Im}\chi_{\alpha\beta}(\mathbf{q},\omega)}{\hslash \omega},
\end{equation}
where the form factors $\mathcal{F}_{\alpha\beta}(\mathbf{q})$ (see Appendix A for details) depend on the local dipolar hyperfine couplings, and $\chi_{\alpha\beta}(\mathbf{q},\omega)$ is the dynamical magnetic susceptibility of the Tm moments. For simplicity we only include the two nearest neighbor and four next-nearest neighbor Tm atoms in the form factors.  Because the Tm system exhibits ferroquadrupolar order at $T_Q$, we assume that the structure of the dynamical susceptibility can be modeled as:
\begin{equation}
    \label{eqn:dynamicalchi}
    \chi_{\alpha\alpha}(\mathbf{q},\omega) =\frac{\chi_{\alpha\alpha}(T)}{\xi^{-2} + f(\mathbf{q}) -i\omega/\Gamma q}
\end{equation}
where $\xi$ is a correlation length, $\Gamma$ is a characteristic fluctuation energy, $f(\mathbf{q}) = q_x^2 + q_y^2 + \eta q_z^2$, $\eta$ is a dimensionless parameter that reflects the tetragonal nature, and $\chi_{\alpha\alpha}(T)$ is the static ($\mathbf{q}=0$) susceptibility. $\xi$ and $\eta$ are unknown parameters, but we compute the temperature and angular dependence using $\xi=2$ and $\eta = 1/2$.  Fig. \ref{Fig:T1inv}(b) shows the expected angular dependence of $T_{1m}^{-1}$ for  $\chi_{cc}/\chi_{aa} = 10$ (red), close to the experimental value, and for   $\chi_{cc}/\chi_{aa} = 0.1$ (blue). The former clearly exhibits a maximum of $T_{1m}^{-1}$ at $\theta=90^{\circ}$, in contrast to our observations. The latter exhibits a shallow minimum at $90^{\circ}$, but the susceptibility anisotropy does not agree with experiment.  Fig. \ref{Fig:T1inv}(c) shows the temperature dependence using the measured values of the static susceptibility. Although there is an overall decrease in $T_{1m}^{-1}$ at lower temperatures, the detailed temperature dependence does not match experiment, and the calculated \slrr\ still exhibits a maximum for $\theta = 90^{\circ}$ at all temperatures.  Despite the complex form factors for the direct dipolar couplings, the expected magnetic fluctuations of the Tm ground state cannot explain the observed increase in \slrr\ as the field rotates out of the plane.

\subsection{Quadrupolar fluctuations}

An alternative explanation is that the \slrrtext\ is dominated  quadrupolar fluctuations rather than magnetic.  The Tm quadrupole moments couple to the EFG at the V site, giving rise to a second nuclear quadrupolar relaxation channel \cite{DioguardiPdoped2015}.  The enhancement of \slrr\ below 20K for $\theta = 90^{\circ}$ may represent the growth of critical fluctuations near $T_Q$.  Note that changing $\theta$ by only 0.25$^{\circ}$ dramatically alters \slrr, which is close to the limit of precision of our goniometer.  Thus it is possible that the enhancement below 20K may vanish or become smaller for better alignment.   In the presence of both magnetic and quadrupolar relaxation, the expression for $\phi(t)$ (Eq. \ref{eqn:relax}) changes, and includes three independent rates: $T_{1m}^{-1}$, $W_{Q1}$ and $W_{Q2}$, where the latter two are associated with $\Delta m = \pm 1$ and $\Delta m = \pm 2$ quadrupolar relaxation. We are unable, however, to independently extract these parameters with sufficient resolution.  Moreover, the line broadening observed in Fig. \ref{Fig:spectra} also means that the magnetization relaxation at the central transition may also include contributions from nearby satellite transitions, further complicating any attempts to extract the independent relaxation channels.  Nevertheless, it is instructive to consider the case where quadrupole fluctuations dominate and magnetic fluctuations can be neglected.

Quadrupolar relaxation is driven by fluctuations of the spherical tensor components of the EFG: $V_{\pm 1} = V_{zx}\pm iV_{zy}$ and $V_{\pm2} = \frac{1}{2}(V_{xx} - V_{yy}) \pm i V_{xy}$, where the $V_{\alpha\beta}$ are the EFG tensor components relative to  the direction of $\mathbf{H}_0$. 
These give rise to nuclear spin relaxation rates: 
\begin{equation}
\label{eqn:wq}
    W_{Q1,Q2} = (eQ/\hbar)^2\int_0^{\infty}\langle V_{+1,2}(\tau)V_{-1,2}(0)\rangle e^{-\omega_L \tau}d\tau
\end{equation}
where $\omega_L$ is the Larmor frequency \cite{suterquadrupolarrelaxation}. The nematic order in this system has $B_{2g}$ symmetry, so  $V_{xx}-V_{yy}\neq 0$, where $z$ corresponds to the $c$-direction and $x$ and $y$ are along the principal axes of the EFG tensor, which are rotated 45$^{\circ}$ relative to the tetragonal $a$-axes. Above $T_Q$ fluctuations of $V_{\pm2}$ should dominate those of $V_{\pm1}$, and as a result we anticipate that $W_{Q1}(\theta =0)$ can be neglected. As the field is rotated towards the plane, the EFG tensor components change, and  the relaxation rates become angular dependent (see Appendix B for details):
\begin{eqnarray}
\label{eqn:quadangle1}
W_{Q2}(\theta)/W_{Q2}(0) &=&\left(\cos ^4\theta+6 \cos ^2\theta +1\right)/8\\
\label{eqn:quadangle2}
W_{Q1}(\theta)/W_{Q2}(0) &=& \sin ^2\theta (\cos (2 \theta )+3)/4.
\end{eqnarray}
These quantities are shown in Fig. \ref{Fig:T1inv}(b) as dashed lines. $W_{Q2}$ exhibits a minimum for $\theta = 90^{\circ}$, whereas $W_{Q1}$ is nearly independent of $\theta$ at this angle. This behavior agrees qualitatively with our observations, but the increases we observe are in fact a much stronger function of angle than expected for quadrupolar relaxation. Rotating $\theta$ by 1-2$^{\circ}$ out of the plane enhances \slrr\ by an order of magnitude, whereas $W_{Q2}$ exhibits only a quadratic minimum at this angle.

\begin{figure}
\centering
\includegraphics[width=\linewidth]{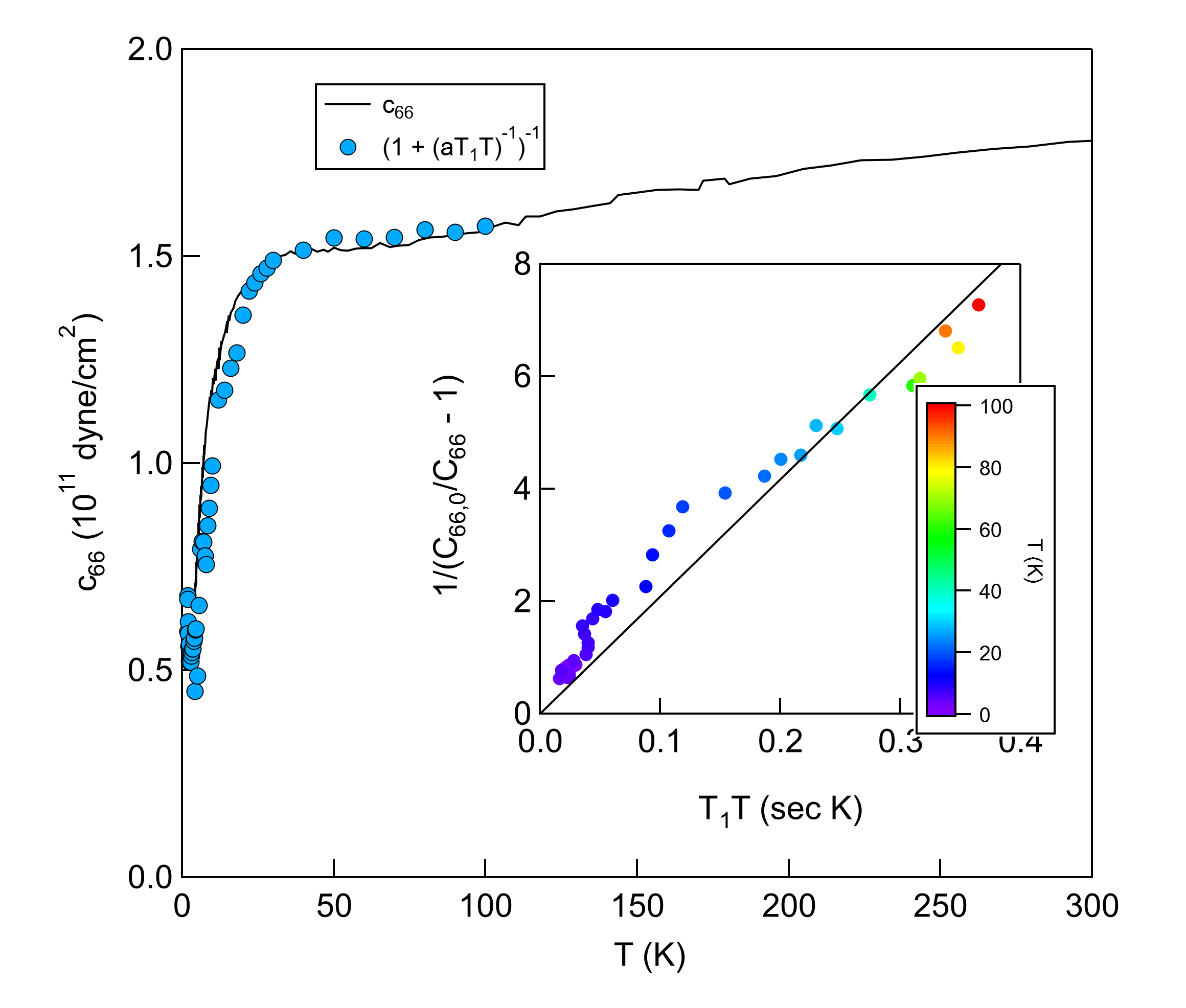}
\caption{\label{Fig:c66} The shear elastic stiffness coefficient $c_{66}$  (solid line, reproduced from \cite{TmVO4CJTE}) and the quantity $1/(1+(aT_1T)^{-1})$ as as a function of temperature.  INSET: $1/(c_{66,0}/c_{66} -1)$ versus $T_1T$, with temperature implicit. The solid black line is the best linear fit, giving 
$a=18.5\pm0.4$ sec$^{-1}$ K$^{-1}$.}
\end{figure}

The interpretation that relaxation is driven by quadrupole fluctuations is supported by comparisons of the temperature dependence of \slrr\ with that of the shear elastic stiffness coefficient, $c_{66}$, which softens with decreasing temperature and vanishes at $T_Q$ \cite{TmVO4CJTE}.  
{This behavior is driven by the nematic susceptibility: $\chi_{nem} = c_{66,0}(1-c_{66,0}/c_{66})/\lambda^2$, where $\lambda$ is the coupling between the lattice and the Tm 4f orbitals, and $c_{66,0}$ is the stiffness coefficient in the absence of the coupling \cite{FernandesPRLnematicT1}. If \slrr\ is also determined by the Tm orbital fluctuations, then $(T_1T)^{-1} \sim \chi_{nem}$ (see Appendix C for details) \cite{DioguardiPdoped2015,Leung1979,Page1983}. We thus expect $T_1T \sim (c_{66,0}/c_{66}-1)^{-1}$, which is demonstrated in Fig. \ref{Fig:c66}. The main panel compares the temperature dependence of $c_{66}$ with the measured \slrr\ values, and the inset shows the scaling between the shear modulus and \slrr\ with temperature as an implicit parameter.  The scaling evident in Fig. \ref{Fig:c66} suggests that the spin lattice relaxation is driven primarily by quadrupole fluctuations, which are reflected in the softening of $c_{66}$. }  


\section{Discussion}

{A slightly different scaling relationship was found in the iron pnictide superconductors via a microscopic model that assumes that the nematicity arises in the magnetic susceptibility, which in turn affects the nuclei through a magnetic hyperfine interaction \cite{FernandesPRLnematicT1}}. In \tmvo, the nematicity arises from the Tm electronic orbitals, and the coupling to the nuclei may be through the quadrupolar interaction.  Moreover, the pnictide model assumed the presence of Landau damping by a Fermi surface of quasiparticles, which is not the case for insulating \tmvo.  
{The relaxation in \tmvo\ must also involve a damping term, but the origin of this term is unknown. The fact that the scaling relationship in Fig. \ref{Fig:T1inv} holds suggests that this damping term is temperature independent.}

Rotating the field away from $90^{\circ}$ can enhance quadrupole fluctuations.  A rotation of $\mathbf{H}_0$ by $4^{\circ}$ corresponds to a field of $0.82$ T along the $c$-axis. This is greater than the critical field of $H_c = 0.52$ T to suppress the long-range nematic order, which naturally enhances fluctuations of both $V_{\pm2}$ and $V_{\pm1}$.  However, these critical fluctuations are not likely to persist to higher temperatures beyond $\sim 10T_Q$, thus are unlikely to be responsible for the large anisotropy observed up to 80K.  An alternative scenario is that the higher CEF levels cannot be ignored.  Indeed, even though the in-plane g-factor of the non-Kramers doublet vanishes in zero applied field, the excited CEF levels can be mixed into the ground state wavefunctions by an in-plane field. As a result, there can be an induced magnetic moment in the plane, which may also contribute to the relaxation \cite{TmVO4CJTE}.

It is likely that the spin lattice relaxation is dominated by both magnetic and quadrupolar fluctuations, however it is difficult to disentangle these two relaxation channels without more detailed measurements of the relaxation at the higher satellite transitions  \cite{suterquadrupolarrelaxation}. However, as illustrated in Fig. \ref{Fig:spectra}, the satellites are magnetically broadened and cannot be well resolved, especially at lower temperatures.  This broadening is due to the  demagnetization field inhomogeneity of our crystal. In principle, it is possible to improve the spectral resolution by removing the sharp edges and corners of the sample and/or operating at lower applied fields, in order to better discern the individual satellite transitions. 

{Nuclear spin lattice relaxation rates have also been studied in both PrAlO$_3$ and CsCuCl$_3$, materials that exhibit structural distortions due to the cooperative Jahn-Teller effect with non-magnetic ground states  \cite{Borsa1978,Corti1981,Rigamonti1984}. In contrast to our observations in \tmvo,  \slrr\ did not exhibit any enhancement above the phase transition in these cases, even though the EFG changed below. On the other hand, unlike \tmvo, the phase transitions in these cases are first order, thus \slrr\ should not reflect any critical slowing down. The \slrrtext\ in the disordered state was analyzed in terms of magnetic (hyperfine) fluctuations, although the nuclei in question ($^{27}$Al, $I=5/2$ and $^{133}$Cs, $I=7/2$) are quadrupolar and should be sensitive to fluctuations of the EFG. }

In summary, we have measured the spectra and relaxation rates in \tmvo\ as a function of field direction oriented perpendicular to the c-axis.  We find that the magnetic shift tensor agrees quantitatively with direct dipolar coupling between the V nuclear moments and the Tm 4f moments.  The spin lattice relaxation rate exhibits a steep minimum for field oriented 90$^{\circ}$ to the $c$ axis, which is inconsistent with purely magnetic fluctuations. We find that $T_1$ scales with the lattice constant for shear strain, $c_{66}$, which softens and vanishes at the nematic transition.  It is likely that both quadrupolar and magnetic fluctuations are present and drive spin lattice relaxation.  However, the origin of the the steep angular dependence of \slrr\ remains an open question.  

\section{Acknowledgments}

We acknowledge helpful discussions with R. Fernandes. Work at UC Davis was supported by the NSF under Grants No. DMR-1807889 and PHY-1852581, as well as the UC Davis Seed Grant program. Crystal growth performed at Stanford University was supported by the Air Force Office of Scientific Research under award number FA9550-20-1-0252. P. M. was partially supported by the Gordon and Betty Moore Foundation Emergent Phenomena in Quantum Systems Initiative through Grant GBMF9068.

\appendix
\section{Magnetic Relaxation Form Factors}

We assume that the dominant hyperfine fields at the V site arise from the two nearest neighbor and four next-nearest Tm moments, whose positions are given in Table \ref{tab:positions}. We define:
\begin{equation}
   \mathcal{A}_{\alpha\beta}(\mathbf{q}) = \sum_i e^{i\mathbf{q}\cdot\mathbf{r}_i} A_{\alpha\beta}^{dip} 
\end{equation}
where $A_{\alpha\beta}^{dip}$ is defined in the main text.  For an applied field $\mathbf{H}_0$ oriented at angles $\theta$ and $\phi$ relative to the crystalline axes, the form factors are \cite{T1formfactorsArsenides}:
\begin{equation}
  \mathcal{F}_{\alpha\beta}(\mathbf{q}) =  \sum\limits_{\epsilon,\delta} \left[R_{x\epsilon}R_{x\delta} + R_{y\epsilon}R_{y\delta} \right]\mathcal{A}_{\epsilon\alpha}(\mathbf{q})\mathcal{A}_{\delta\beta}(-\mathbf{q}),  
\end{equation}
where the $R_{\alpha\beta}$ are elements of the 3D rotation matrix:
\begin{widetext}
\begin{equation}
  \mathbb{R} = \left(
\begin{array}{ccc}
 \cos \theta  \cos ^2\phi+\sin ^2\phi & \cos \theta \cos \phi\sin \phi-\cos \phi \sin \phi &
   \cos \phi  \sin \theta \\
 \cos \theta  \cos \phi \sin \phi-\cos\phi  \sin \phi  & \cos ^2\phi +\cos \theta  \sin ^2\phi &
   \sin \theta \sin \phi \\
 -\cos \phi \sin \theta  & -\sin \theta \sin \phi  & \cos \theta
\end{array}
\right).
\end{equation}
\end{widetext}

\begin{table}[t]
\caption{\label{tab:positions}Position vectors for six nearest neighbor Tm sites to V, in spherical coordinates.}
\begin{ruledtabular}
\begin{tabular}{lccc}
$\mathbf{r}_i$ & $r$ (\AA) & $\theta$ ($^{\circ}$) & $\phi$ ($^{\circ}$)\\
\hline
1 & 3.13030 & 0 & 0 \\
2 & 3.13030 & 180 & 0 \\
3 & 3.86654 & 66.1218 & 0 \\
4 & 3.86654 & 113.8782 & 90 \\
5 & 3.86654 & 66.1218 & 180 \\
6 & 3.86654 & 113.8782 & 270
\end{tabular}
\end{ruledtabular}
\end{table}

\section{Quadrupolar Relaxation Anisotropy}

Equation \ref{eqn:wq} gives the expression for quadrupolar relaxation in terms of the spherical tensor components of the EFG tensor.  The quadrupolar interaction is only on-site, so there are no form factors.  However, the EFG tensor must be rotated properly as the field direction changes.    
Under a rotation the tensor operators $V_{m}(\tau)$ transform as:
\begin{equation}
    V'_{m}(\tau) = \sum_{m'}D_{mm'}^{(2)}V_{m'}(\tau)
\end{equation}
where 
\begin{equation}
  D_{mm'}^{(l)}(\alpha,\beta,\gamma) = e^{-i m\alpha} d^l_{mm'} (\beta) e^{-i m'\gamma},
\end{equation}
are the Wigner $D$ matrices, and the Euler angles are $(\alpha = \phi,\beta=\theta,\gamma=0)$. The correlation functions $\langle V_{m}(\tau)V_{-m}(0)\rangle$ are thus given by:
\begin{widetext}
\begin{equation}
  \langle V'_{m}(\tau)V'_{-m}(0)\rangle = \sum_{m',m''}D_{2m'}^{(2)}(\phi,\theta)D_{-2m''}^{(2)}(\phi,\theta)\langle V_{m'}(\tau)V_{m''}(0)\rangle.
\end{equation}
\end{widetext}
We assume that $\langle V_{m}(\tau)V_{m'}(0)\rangle =0$ for all $m,m'$ except for $m = -m' = 2$ and $m = m' = 1$.  Moreover, we assume that $\langle V_{2}(\tau)V_{-2}(0)\rangle \gg \langle V_{1}(\tau)V_{-1}(0)\rangle$, since $\langle V_{\pm2}\rangle \neq 0$ and $\langle V_{\pm1} \rangle =0$ in the nematic phase.  We thus expect $W_{Q1}(\theta =0)\approx 0$, and:
\begin{eqnarray}
W_{Q2}(\theta)/W_{Q2}(0) &=&\left(\cos ^4\theta+6 \cos ^2\theta +1\right)/8\\
W_{Q1}(\theta)/W_{Q2}(0) &=& \sin ^2\theta (\cos (2 \theta )+3)/4.
\end{eqnarray}
as given in the main text in Eqs. \ref{eqn:quadangle1} and Eqs. \ref{eqn:quadangle2}.

\section{Relaxation driven by nematic fluctuations}
We note that Eq. \ref{eqn:wq} can be expressed in terms of the dynamical nematic susceptibility \cite{DioguardiPdoped2015}:
\begin{equation}
W_{Q2}(0)=\left(\frac{eQ}{\hbar}\right)^{2}k_{B}T\lim_{\omega\rightarrow0}\sum\limits _{\mathbf{q}}\frac{\mathrm{Im}\chi_{nem}(\mathbf{q},\omega)}{\hslash\omega}.\label{eqn:Moriya_nematic}
\end{equation}
The dynamical susceptibility can be expressed phenomenologically as: 
$\chi_{nem}(\mathbf{q},\omega) = {\chi_{nem}}({1-i\omega/\omega_n})^{-1}$, where
where $\chi_{nem}$ is the static nematic susceptibility and $\omega_n$ is a damping term \cite{Leung1979,Page1983}.  In this case $W_{Q2}(0) =\left({eQ}\right)^{2}k_{B}T \chi_{nem}/\hbar^2\omega_n$. 
\bibliography{TmVO4NMRbibliography}

\end{document}